\documentclass[a4paper,11pt]{article}
\usepackage{jinstpub} 
\usepackage{lineno}
\usepackage{siunitx}
\usepackage{booktabs}
\usepackage{amsmath,amsfonts}
\usepackage{mathastext}
\usepackage{orcidlink}
\usepackage[font=normalsize,labelfont=sf,textfont=sf]{subfig}
\providecommand{\RSfigtxt}{Fig.\,}

\DeclareSIUnit \rms {\text{RMS}} 

\newcommand{\tbt}{\qtyproduct{2x2}{} }
\title{Characterization of a low-power 3D photon-to-digital converter readout improved for system integration in meter-scale applications}

\author[a,1]{O.~Lepage\orcidlink{0009-0003-1645-1578}\note{Corresponding author.}}
\author[a]{, T.~Rossignol\orcidlink{0000-0002-9270-9193}}
\author[a]{, N.~Roy\orcidlink{0000-0002-4718-210X}}
\author[a]{, G.~Lessard\orcidlink{0009-0006-7358-9890}}
\author[a]{, F.~Vachon\orcidlink{0000-0002-6976-6624}}
\author[b]{, L.~Fabris\orcidlink{0000-0001-5605-5615}}
\author[a]{, S.A.~Charlebois\orcidlink{0000-0001-7857-5056}}
\author[a]{, J.-F.~Pratte\orcidlink{0000-0002-8327-3842}}
\affiliation[a]{Interdisciplinary Institute for Technological Innovation and Department of Electrical and Computer Engineering, Université de Sherbrooke,\\3000 Boulevard de l’Université, Sherbrooke, Canada}
\affiliation[b]{Oak Ridge National Laboratory\\1 Bethel Valley Rd \#5200, Oak Ridge, TN 37830, USA.}

\emailAdd{Grams3D-Info@usherbrooke.ca}

\abstract{Digital silicon photomultipliers (dSiPM) are arrays of single photon avalanche diodes (SPADs) where each SPAD has its own electronic readout. 
To maximize the photodetection area, we developed a readout integrated circuit (ROIC) 3D-integrated to a custom-designed SPAD layer fabricated at Teledyne Dalsa (Bromont, Canada) to form a photon-to-digital converter (PDC). 
This paper presents an improved version of a previously demonstrated ROIC, fabricated in TSMC \SI{180}{\nano\meter} technology to manage a \qtyproduct{64x64}{} pixel architecture.
This ROIC has different outputs, such as a flag output that produces a pulse whenever one of the pixels triggers and a digital sum that samples the amount of triggered SPADs.
Improvements lead to a reduction of the timing jitter on the flag output from \SI{72.0}{\pico\second\rms} to \SI{22.6}{\pico\second\rms} through optimized H-tree design. 
A tunable hold-off circuit provides adjustable dead time from \SI{32}{\nano\second} to \SI{18}{\micro\second}, addressing both afterpulsing mitigation and SPAD-to-SPAD variations due to process defects.
Power consumption is lowered by \SI{30}{\percent} at trigger rates exceeding \SI{10}{\kilo\hertz} through digital logic optimization and clock gating. 
These measurements validated that the ROIC is ready for 3D integration into a PDC for systems in medical imaging, particle physics, and quantum sciences.}


\begin{document}
\maketitle
\flushbottom

\section{Introduction}\label{sec:intro}
Silicon photomultipliers (SiPM) are used in a wide range of applications, such as medical imaging, meter-scale particle experiments, and quantum sciences \cite{herwegAnalyzingTimeDistribution2025, acerbiCompactHandheldSiPMBased2025, delgadoComparativeStudySiPM2024, aalsethDarkSide20k20Tonne2018, zhaoFeasibilityHighresolutionPET2019, radichNeutronSpectrometerNeutron2024, carrierMultiPixelPhotontoDigitalConverter2023}.
These photodetectors output a charge quantity proportional to the number of triggered single photon avalanche diodes (SPADs) \cite{bragaMiniSiPMArrayPET2011, CALO201957}.
The sum of these current pulses passes through a transimpedance amplifier followed by signal shaping, analog-to-digital conversion, and signal processing to determine the number of detected photons.
Taking advantage of the boolean nature of the SPAD, digital SiPMs incorporate a network of transistors to individually quench each SPAD, delivering a digital pulse for each photon detected \cite{frachDigitalSiliconPhotomultiplier2009}.
This paper presents an improved version of the photon-to-digital converter readout integrated circuit (ROIC) presented in \cite{rossignol3DPhotontoDigitalConverter2024}, with an emphasis on low-power and versatility.
This PDC readout aims for static microwatt-range power operation, sub-\SI{100}{\pico\second} timing precision, and variable sampling rates ranging from \SI{10}{\nano\second} up to \SI{10}{\micro\second} to accommodate a large spectrum of application requirements.
In addition, this ROIC is designed for and used to produce PDCs by 3D integration with a tailored SPAD layer made in a dedicated process at Teledyne DALSA (Bromont, Canada) \cite{parentDesignVerticallyIntegrated2022a, pratte3DPhotonToDigitalConverter2021a, parentWaferLevelCharacterizationMonitoring2024a}.
Heterogeneous vertical integration ensures efficient area utilization and enables independent selection of the optimal technologies for the SPAD and the electronics, individually \cite{pratte3DPhotonToDigitalConverter2021a}.
The ROIC is also designed with built-in debugging features, such as reconfigurable IOs and a triggering system.
This trigger enables characterization of the quench circuit independently of SPAD triggering, reducing test time and facilitating known-good-die validation for meter-scale systems.
This article focuses on the ROIC itself independently of the 3D integration and the SPAD layer, keeping in mind the system integration for meter-scale systems.

The rest of this paper is organized as follows:
Section~\ref{sec:arch} outlines the PDC ROIC architecture and meter-scale integration;
Section~\ref{sec:materialsMethods} describes the characterization platforms;
Section~\ref{sec:flag} details event flag timing resolution and electronic jitter;
Section~\ref{sec:mono} presents the adjustable hold-off monostable performance;
and Section~\ref{sec:power} covers power consumption analysis.

\section{Architecture}\label{sec:arch}
\subsection{Overview}\label{subsec:over}
\begin{figure}[!t]
    \centering
    \includegraphics[width=0.8\linewidth]{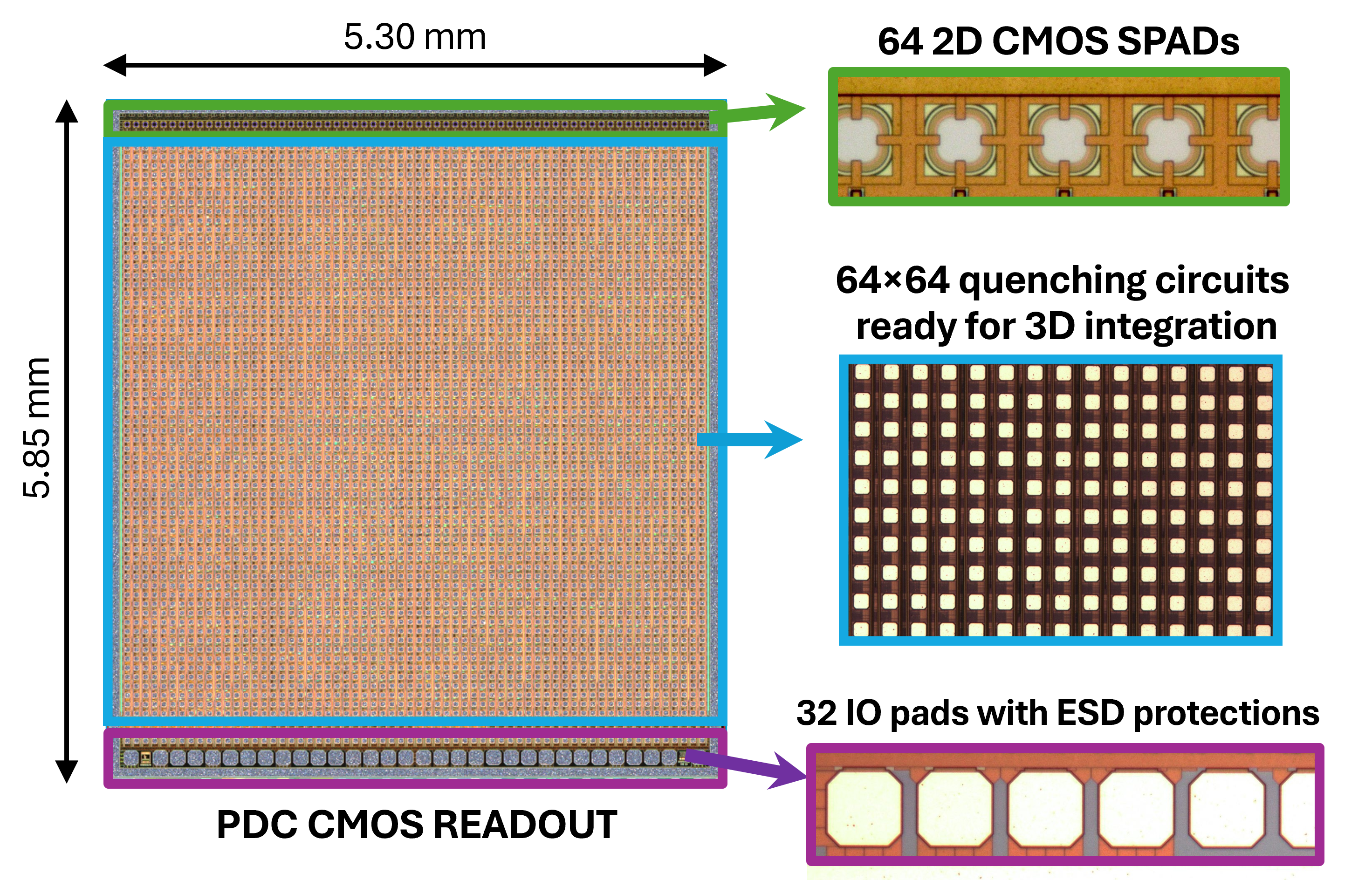}
    \caption{Top view of the PDC ROIC. From bottom to top are the 32 IO pads, the array of \qtyproduct{64x64}{} quenching circuits (QC) surrounded by the digital logic, and finally the row of 64 CMOS SPADs connected to the last row of QCs.}
    \label{fig:arch}
\end{figure}

The ROIC is designed in the TSMC \SI{180}{\nano\meter} process to interface with a vertically bonded layer of \qtyproduct{64x64}{} SPADs.
The layout of the PDC ROIC is divided into 3 sections: the IO pads, the array, and the built-in CMOS SPAD, as shown in \RSfigtxt\ref{fig:arch}.
The lower portion includes the 32 pads for the supplies, the configuration, and the digital communication.
The array includes \qtyproduct{64x64}{} quenching circuits (QC), each incorporating an interconnect pad.
The 4096 pixels cover a \qtyproduct{5x5}{\milli\meter} area with a pitch of \SI{78}{\micro\meter}.
The 64 CMOS SPADs are situated in the upper portion of the layout so they only connect to the last row of QC. 
These SPADs are designed to enable independent ROIC testing prior to 3D integration; however, this article presents only electrical trigger measurements and does not utilize these CMOS SPADs. 
Once 3D bonded, the CMOS SPADs are no longer biased, hence disabled.

\subsection{Quenching circuit}\label{subsec:quench}
\begin{figure*}[!t]
    \centering
    \subfloat[\label{fig:PDC_QC_top}]{%
        \includegraphics[width=0.9\textwidth]{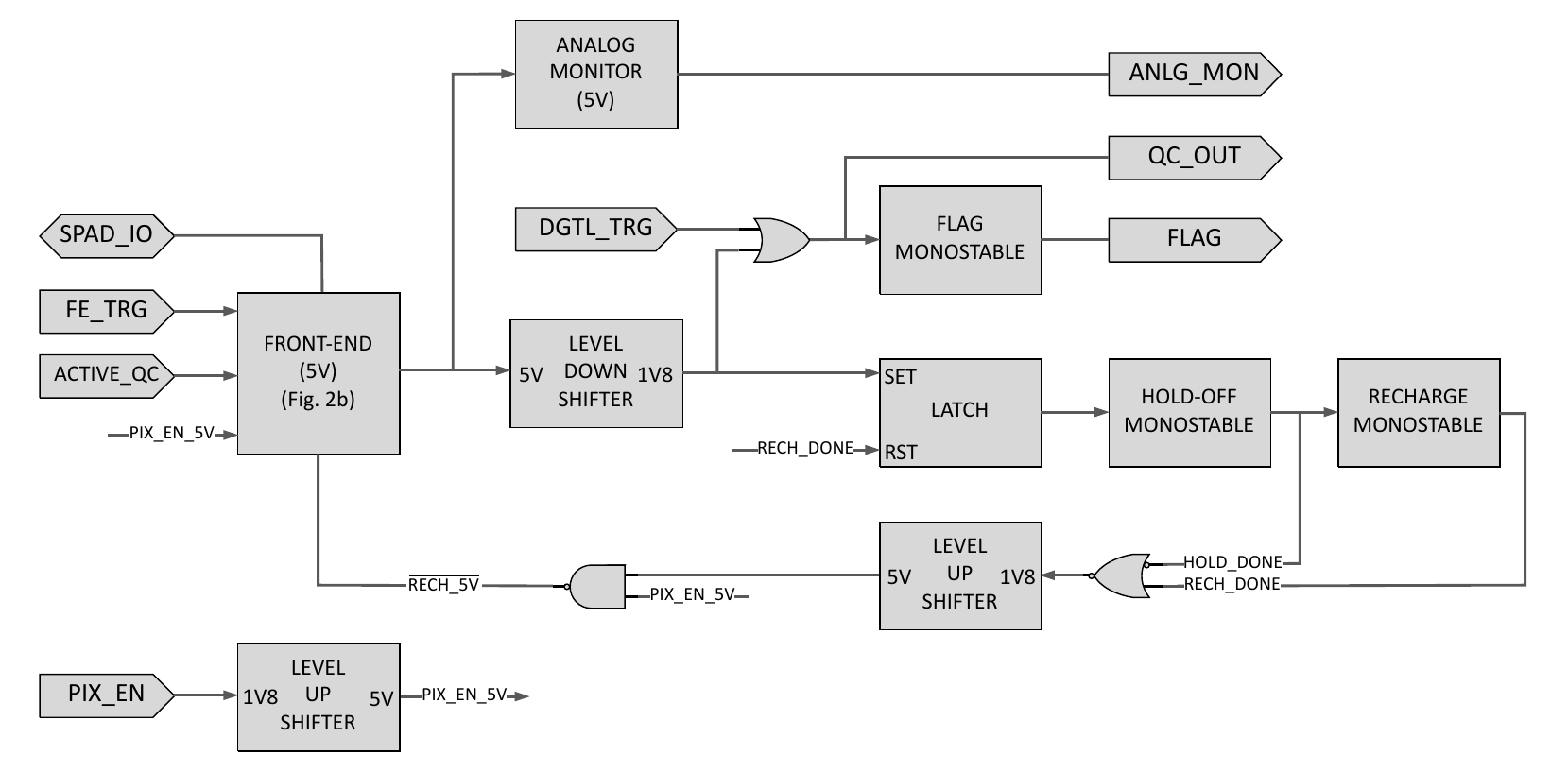}
    }\\
    \subfloat[\label{fig:PDC_QC_FE}]{%
        \includegraphics[width=0.9\textwidth]{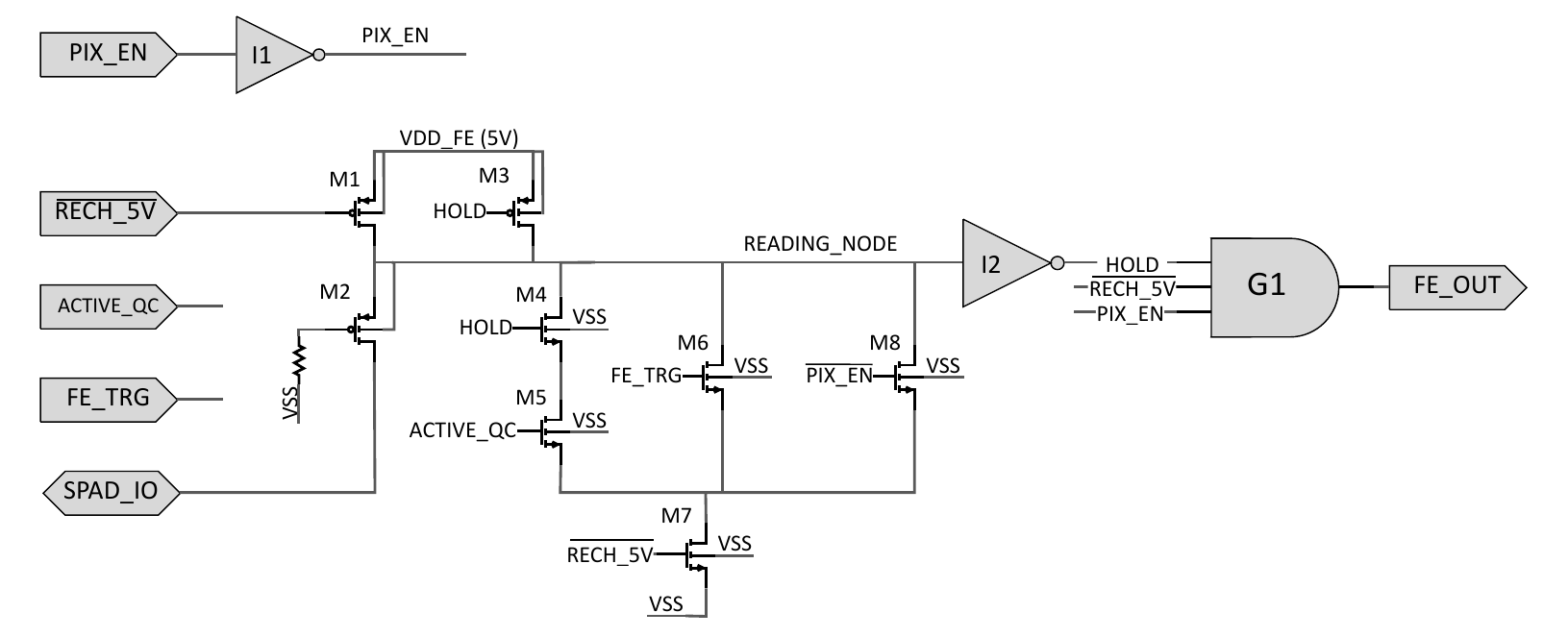}
    }
    \caption{(a) Block diagram of the quenching circuit (QC) of the PDC. The circuit includes the FE but also level-down and level-up shifters, a latch, monostables, and an analog monitor. The QC and all its modules are asynchronous. (b) Front-end transistors of the quenching circuit. SPAD\_IO is connected to the cathode of the SPAD through a 3D bonding pad. The FE is either triggered by the SPAD\_IO or the internal trigger signal FE\_TRG. ACTIVE\_QC enables the active quenching circuit. PIX\_EN enables the QC. FE\_OUT is the output of the front-end circuit, which is then shifted down to 1.8V to be compatible with the digital logic (QC\_OUT).}
\end{figure*}

\RSfigtxt\ref{fig:PDC_QC_top} presents the QC block diagram.
The front-end (FE) block implements \SI{5}{\volt} transistors to interface with the SPAD and produces a digital pulse to indicate detection.
The analog monitor block implements a precise programmable current source to mimic the behavior of an analog SiPM.
A level-down shifter lowers the FE output voltage to \SI{1.8}{\volt} for the core circuitry.
The latch then locks the event signal, and an adjustable monostable sets the hold-off time before the recharge.
A second programmable monostable sets the duration for the SPAD recharge time. 
The event signal is used for both the QC output (QC\_OUT) and the flag output (FLAG).
Each QC is individually connected to an interconnect pad of \qtyproduct{40x40}{\micro\meter}.

\subsubsection{Front-end}
\RSfigtxt\ref{fig:PDC_QC_FE} shows the front-end circuit of the QC.
The SPAD cathode is connected to the SPAD\_IO net.
A cascode (M2) allows increasing the maximum SPAD excess voltage up to \SI{10}{\volt} while keeping the reading node (READING\_NODE, RN) voltage between \SI{0}{\volt} and \SI{5}{\volt}.
While waiting for an event, M3 is closed to keep RN at \SI{5}{\volt}.
When the SPAD fires (or triggers), the RN voltage drops from \SI{5}{\volt} towards \SI{0}{\volt}.
Once RN passes the inverter threshold, M3 opens.
This inverter I2 has been chosen over a voltage comparator to reduce static and dynamic power consumption \cite{noletQuenchingCircuitDiscriminator2023}.
If the active quench feature is enabled (M5 closed), M4 creates a positive feedback to keep RN at \SI{0}{\volt} for the complete duration of the hold-off period.
After the hold-off duration, M1 initiates the recharge of the SPAD while M7 opens to prevent a feedthrough via M4/M5.
The SPAD is then ready for another cycle.

To disable the QC, inverter I1 and M8 keep RN at \SI{0}{\volt}, maintaining M2 open and removing excess voltage from the SPAD.
AND-gate G1 ensures that no signal is output from a disabled QC and keeps the output low during the recharge cycle.
M6, driven by the trigger tree, is used to test the front-end by forcing RN to \SI{0}{\volt} to produce an event-like signal.

\subsubsection{Hold-off, recharge, and flag monostables}
By design, the hold-off duration ranges from \SI{15}{\nano\second} to \SI{18}{\micro\second}.
Three coarse settings adjust the monostable capacitance, while fine settings tune the charging current.
The recharge duration is similarly programmable (\SI{2}{\nano\second} to \SI{60}{\nano\second}) to account for different SPAD sizes and technologies.
For the flag output, shorter pulses improve event discrimination, but a minimum pulse width is needed for external readability.
This duration is tunable from \SI{2}{\nano\second} to \SI{60}{\nano\second}.
These three monostables provide critical flexibility: dark count rate and afterpulse probability vary with temperature and junction capacitance \cite{acerbiNUVVUVSensitive2023, collazuolStudiesSiliconPhotomultipliers2011, otteCharacterizationThreeHigh2017}, making the programmable hold-off, recharge, and flag durations essential for optimal trade-offs between afterpulse suppression and photosensitive window \cite{vachonMeasuringCountRates2021}.

\subsection{Event flag}\label{subsec:flag}
The event flag indicates the arrival of one or more photons by combining the flag pulses of each quenching circuit.
Signals from each QC feed into an H-tree of OR gates with matched routing to minimize delay variation between pixels, compensating for the unknown address of the triggered QC in this ROIC architecture.
The same approach was used in the first version \cite{rossignol3DPhotontoDigitalConverter2024}; however, the routing of the H-tree has been further length-matched in this version. 
Also, extra decoupling cells have been placed near each OR gate to reduce supply variations.
This PDC output enables both single-photon timing via external TDC and multi-PDC coincidence detection on an integrated tile.

\subsection{Digital sum}\label{subsec:sum}
The output states of the 4096 QCs are synchronized during the rising edge of an external clock to perform a fully digital summation, which provides the precise number of pixels triggered within a time interval.
This approach provides a dynamic range of 4096 without losing simultaneous pixel hits while maintaining single-photon precision at higher counts.

The digital sum is sampled by the external clock with a minimum period of \SI{10}{\nano\second}, but this period can be dynamically adjusted for longer integration times.
For example, a \SI{10}{\nano\second} clock captures prompt photons from scintillator decay, while a \SI{50}{\nano\second} clock later samples the event tail.
This clock-on-demand architecture eliminates the need for a permanent on-chip clock, which means the PDC can operate passively until the external clock triggers acquisition.

An internal FIFO with 128 bins buffers the results for later transmission.
Data transmission shares the external clock with the digital sum operation.
Since both operations use the same clock, transmission requires additional cycles per sample, limiting output bandwidth relative to the sample generation rate.

\subsection{Improvements over first readout version}\label{subsec:imp}
The targeted use for this ROIC is its integration in meter-scale systems with low-power specifications.
To this end, clock gating cells have been implemented to let the clock propagate only to the required circuits when needed, hence reducing the dynamic power consumption.
For example, if an acquisition is not initiated, the clock doesn't reach the 4096 synchronizers, which are connected to each QC (QC\_OUT in \RSfigtxt\ref{fig:PDC_QC_top}).

In the previous version, the trigger tree delay non-uniformities limited our ability to measure the flag timing jitter and uniformity.
In this version, the trigger tree is distributed to each pixel following an H-tree structure with an optimization for timing measurements to reduce its contribution to the measurements on the flag tree.
This trigger tree can also trigger either the front-end (FE\_TRG, M6 in \RSfigtxt\ref{fig:PDC_QC_top}) or only the digital logic (DGTL\_TRG, flag, and sum) to help isolate the response of each circuit.

Another improvement over the previous version is the reconfigurable inputs/outputs.
Since each PDC requires dedicated IOs to interface directly with an external system, called the tile controller, minimizing this count reduces the wiring routed to each PDC, thereby decreasing interconnect congestion.
The current version of the ROIC features three communication IOs (CLK, FLAG, and DATA) that require individual connections to the tile controller, plus two configuration links that can be daisy-chained across multiple PDCs, compared to ten connections in the previous version.
The CLK input can be repurposed as a chip select to use with the configuration bus or as a trigger tree input. 
Regarding the outputs, the default functions are the event flag (FLAG) and digital sum (DATA), but they can be reconfigured to multiple functionalities such as a sum threshold flag, configuration/command validation, and trigger output. 
The minimalist I/O design reduces routing complexity to the tile controller, while support for multiple configuration bus schemes (daisy chain, fanout/star, multidrop, point-to-point) ensures architectural flexibility across diverse applications.

\subsection{System integration}\label{subsec:sys}
While an application could use a single PDC, our goal is to provide a way to integrate them into larger area systems.
We currently consider implementations with 32 to 64 ROICs in an array to build a photodetection module (PDM).
These PDMs would be tiled together to create the desired photodetection surface from the \si{\centi\meter\squared}~to~\si{\meter\squared}.
Moreover, this PDM requires a device to manage the configuration of the PDCs, initiate the acquisition and the readout of the digital sum, and acquire precise timestamps of the flags.
The tile controller is taking this role.
This tile controller  manages the configuration, the acquisition, and the extraction of the data by providing a dynamic clock to each PDC.
It also contains functionalities such as time-to-digital converters (TDC), a programmable state machine adaptable to a diversity of applications, a dark count rate (DCR) estimator \cite{vachonMeasuringCountRates2021}, and a DCR filter based on PDC coincidences.
While the end goal is to use an integrated circuit for this task, the measurements presented in this work use an FPGA development board to grant more flexibility during development.

\section{System setup}\label{sec:materialsMethods}
\begin{figure}[!t]
    \centering 
    \includegraphics[width=0.8\columnwidth,scale=1, trim={5 5 5 5}, clip]{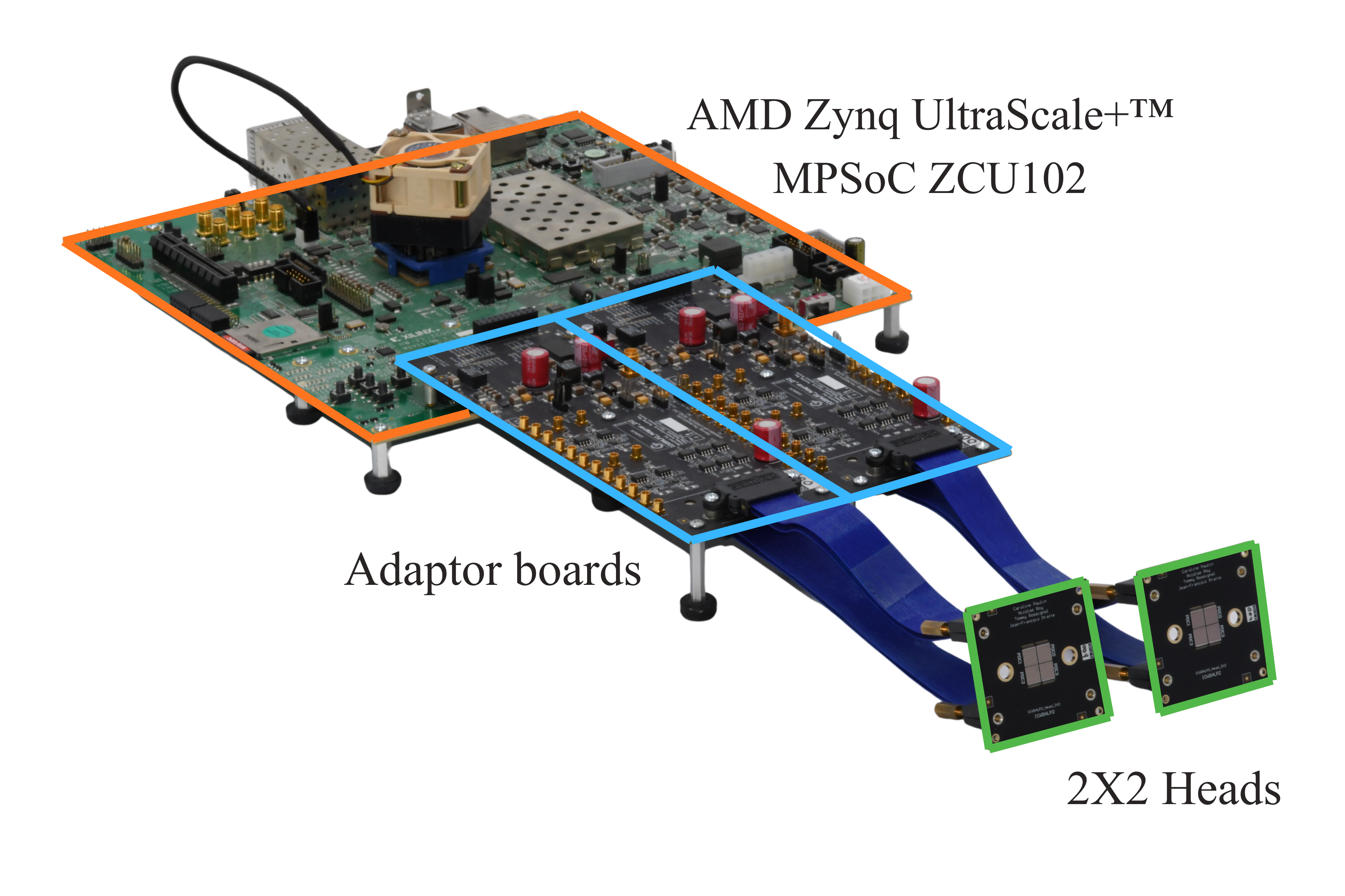} 
    \caption{Measurement setup with two head boards. Each head has \tbt ROIC wirebonded on a single PCB. They are connected to their own adaptor board, interfacing the tile controller. The tile controller is implemented on an AMD Zynq UltraScale+ MPSoC ZCU102 evaluation board.}
    \label{fig:2X2_setup}
\end{figure}

The experimental setup (\RSfigtxt\ref{fig:2X2_setup}) is based on the AMD Zynq UltraScale+ MPSoC ZCU102 evaluation board.
The FPGA acts as the tile controller, and the Zynq part is used as a user interface to communicate with the tile controller for debug purposes.
This development board connects to a custom adaptor board that includes buffers, voltage regulators, and output connectors to monitor with external equipment (e.g., oscilloscope).
Four PDC readouts are arranged in a \tbt tile and wirebonded to the head board. 
This head board has no active component, only decoupling capacitors.
All following results are available in this dataset \cite{lepageDataAndCode}.
\section{Event flag timing precision measurements}\label{sec:flag}
\subsection{Material and methods}\label{subsec:met_time}
\begin{figure}[!t]
    \centering 
    \includegraphics[width=0.75\columnwidth]{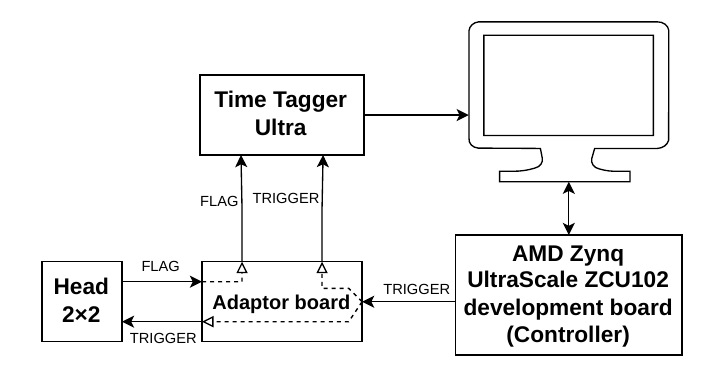} 
    \caption{Block diagram of the setup to measure the flag timing precision. The tile controller sends the trigger signals to the adaptor board, which reroutes it to both the PDCs and the Time Tagger. The latter measures the delay between the trigger and the flag output to build a histogram.}
    \label{fig:block_flag}
\end{figure}

The flag output timing precision measurements rely on a multichannel TDC from Swabian Instruments (Time Tagger Ultra) with \SI{8}{\pico\second\rms} timing precision (\RSfigtxt\ref{fig:block_flag}).
Each pixel of the PDC is enabled one at a time in a random sequence to avoid geometric patterns that could arise from specific effects (e.g., temperature variation).
For each pixel, the FPGA sends multiple triggers to both the PDCs and the Time Tagger, which measures the delay between the trigger and the flag output to build a histogram. 
With one histogram per pixel, the jitter and the average delay can be extracted and mapped to the geometry of the PDC. 
The average delay is an absolute measurement that includes the cable length and the buffers on the PCB.
Since the relevant result is the pixel-to-pixel variation, this contribution is removed. 
Both results (the jitter and the pixel-to-pixel variation) are combined to form the global flag timing precision.
\subsection{Results}\label{subsec:res_time}

\begin{figure}[!t]
    \centering
    \includegraphics[width=0.8\columnwidth]{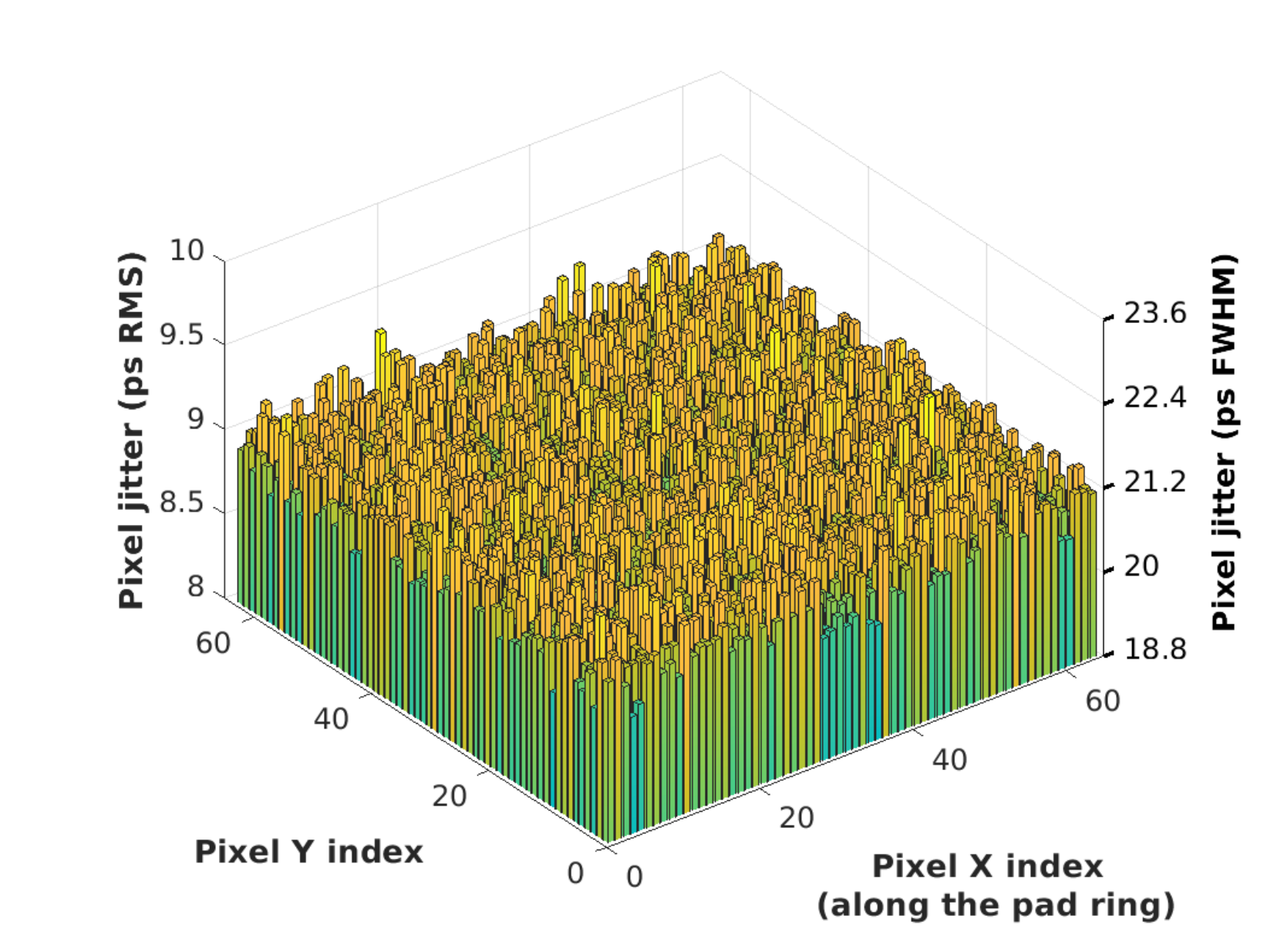} 
    \caption{Pixel map of one ROIC of the per-pixel electronic jitter, including the trigger tree, the QC electronics, the flag tree, and the ROIC output buffer. The measured jitter is \SI{8.93}{\pico\second\rms} in average.}
    \label{fig:FWHM_map}
\end{figure}

\RSfigtxt\ref{fig:FWHM_map} shows the pixel jitter map for one ROIC.
The global electronic jitter over this PDC averages \SI{8.93}{\pico\second\rms}.
The measurements include the PDC contributions (QC electronics, trigger tree, flag tree, and PDC output buffers) and the PCB contribution (fanout buffers).

\begin{figure*}[!t]
    \centering
    \subfloat[Simulations]{\includegraphics[width=0.45\columnwidth]{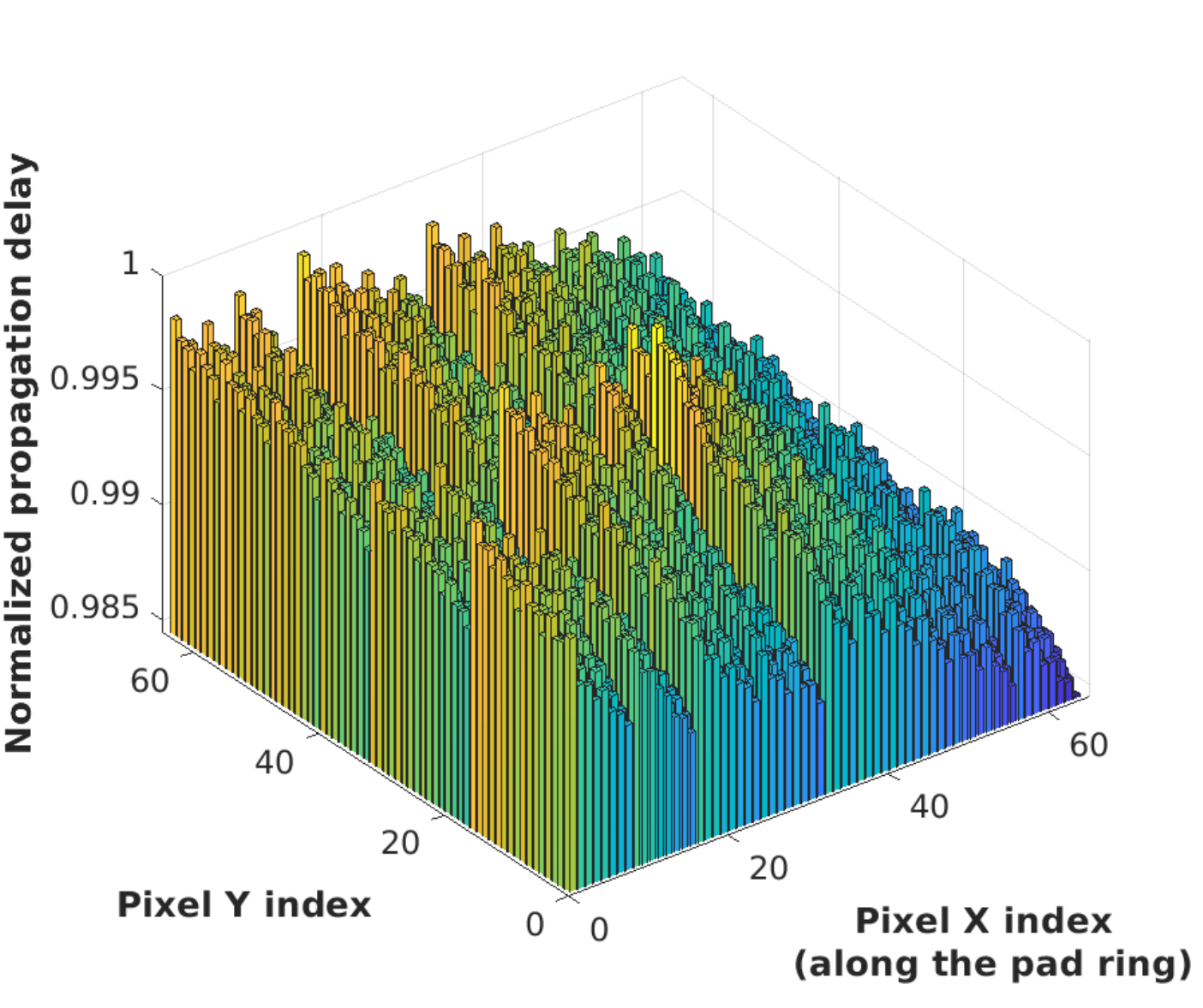}} 
    \subfloat[Measurements]{\includegraphics[width=0.45\columnwidth]{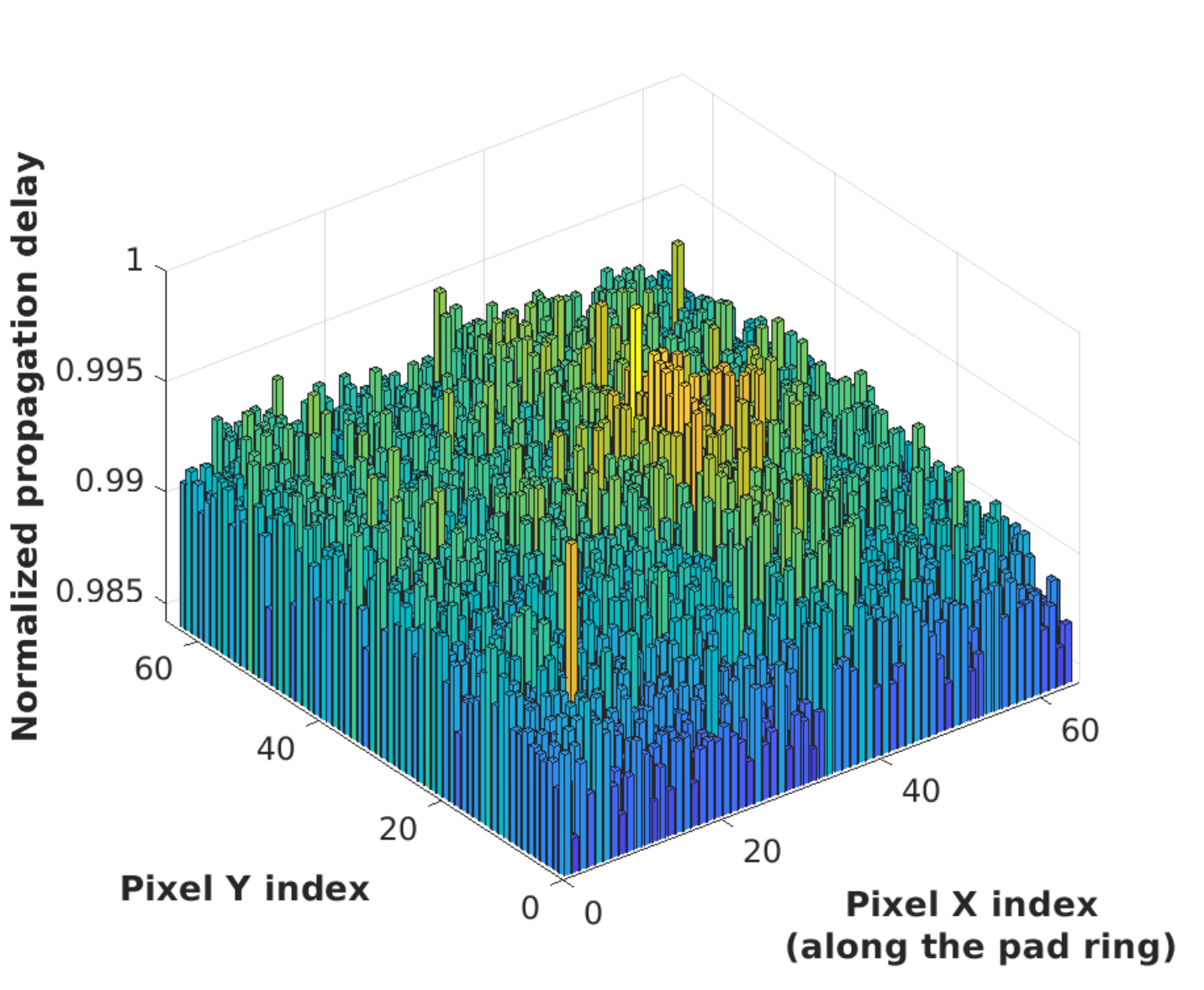}} 
    \caption{Normalized propagation delay maps (delay divided by maximum) showing routing non-uniformity across the pixel array. Color coding follows the z value from blue to yellow (increasing).}
    \label{fig:PROP_map_sim}
\end{figure*}

\RSfigtxt\ref{fig:PROP_map_sim} shows pixel maps of the simulated and measured propagation delay across the array.
Each delay value is normalized by the maximum delay in the distribution to highlight routing non-uniformities.
An increasing gradient from the bottom-right corner to the top-left corner is observable in both maps but more noticeable in the simulation.
This gradient reflects the H-tree routing geometry, where pixels passing through the gate's fastest input branch (bottom right) exhibit shorter propagation delays than those using the slowest branch (top left).
The center of the array has a higher interconnection density that increases the parasitics, which leads to the observed longer propagation delay.
\begin{figure}[!t]
    \centering
    \includegraphics[width=0.8\linewidth]{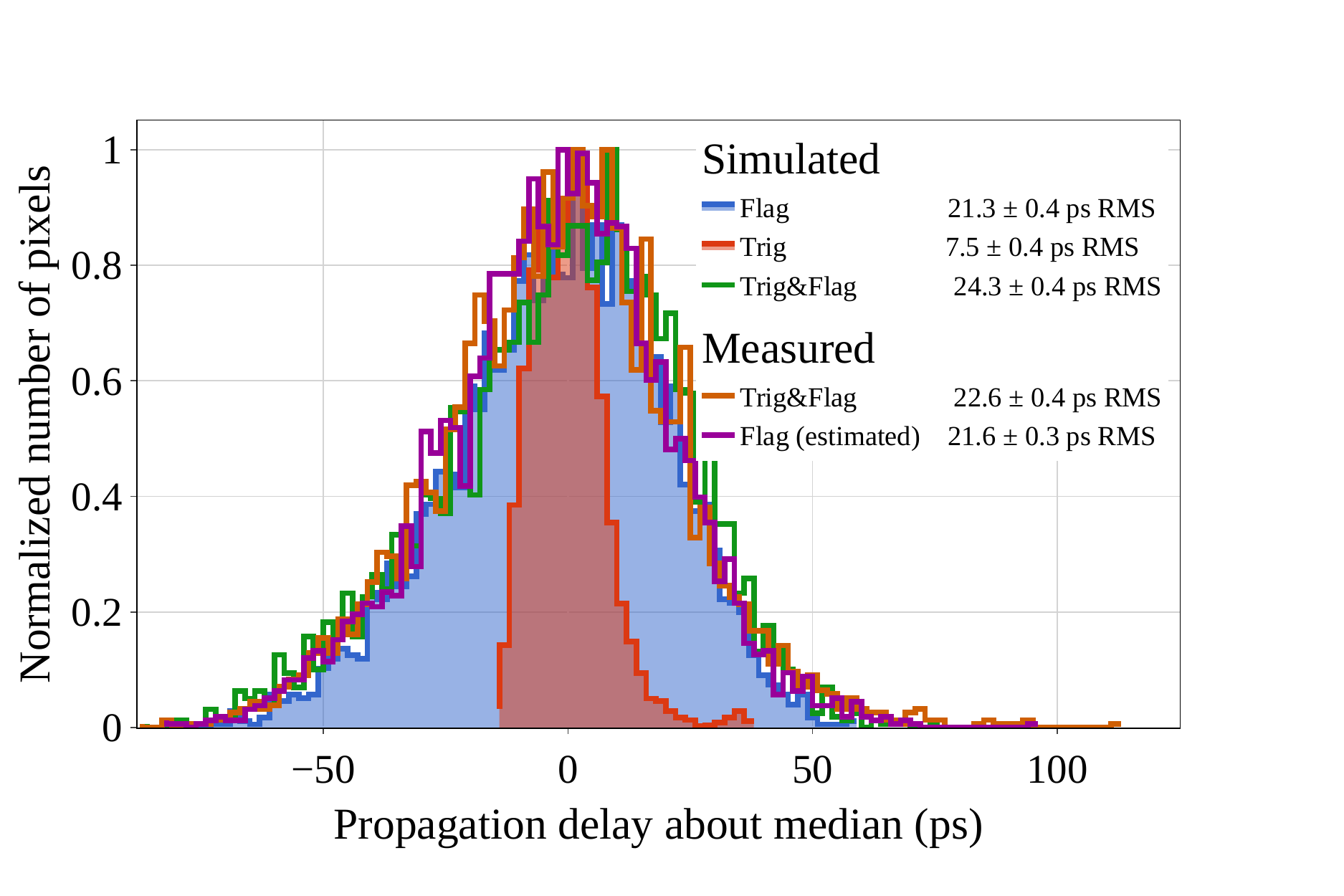} 
    \caption{Propagation delay distribution across pixels, about the median value. Measured delays are compared against simulated flag and trigger propagation delays, revealing pixel-to-pixel skew relative to the array median.}
    \label{fig:ND_FE}
\end{figure}

\RSfigtxt\ref{fig:ND_FE} presents the measured propagation delays about the median, alongside simulated flag and trigger delays.
Since pixel address information is not available in this architecture, pixel-to-pixel propagation skew directly translates to system-level timing jitter.
The skew distribution over the ROIC is \SI{\sim22.6}{\pico\second\rms}.

\subsection{Discussion}\label{subsec:dis_time}
The electronic jitter of this implementation is slightly higher than in \cite{rossignol3DPhotontoDigitalConverter2024}. 
This might be attributable to changing the digital library cell to an area-optimized version.
On the other hand, the skew between pixels was reduced from \SI{89.0}{\pico\second\rms} in the first revision to \SI{22.6}{\pico\second\rms} in this version.
This improvement comes mostly from the trigger tree, which was redesigned using a symmetric H-tree distribution network. 
In the previous implementation, the trigger signal distribution relied on standard automatic routing, resulting in larger path mismatches across the array and increased timing skew.
The shape of the distribution of the normalized propagation delays also changed with this improvement to follow more closely a Gaussian distribution compared to the first version.
Furthermore, the distribution follows more accurately the simulated flag and trigger, as shown in \RSfigtxt\ref{fig:ND_FE}.

Table \ref{tab:timing} presents the results of the 4 PDCs tested for this revision, compared to the same measurements made on the first revision.

\begin{table}[!t]
    \centering
    \renewcommand{\arraystretch}{1.2}
    \begin{tabular}{llcccc||c}
    \toprule
                                    &    & PDC0 & PDC1 & PDC2 & PDC3 & \cite{rossignol3DPhotontoDigitalConverter2024} \\
    \midrule
    \midrule
        &Jitter (\si{\pico\second})      & 8.93 & 12.2 & 8.17 & 10.81 & 7.70 \\
        RMS & Skew (\si{\pico\second})   & 22.6 & 26.1 & 23.6 & 30.2 & 89.0 \\
        &Skew\&Jitter (\si{\pico\second}) & 24.8 & 35.1 & 25.2 & 32.1 & N/A \\
    \midrule
        &Jitter (\si{\pico\second})      & 19.4 & 25.7 & 24.3 & 24.6 & N/A \\
        FWHM & Skew (\si{\pico\second})  & 54.0 & 61.0 & 55.0 & 71.0 & N/A \\
        &Skew\&Jitter (\si{\pico\second}) & 58.0 & 88.0 & 56.0 & 74.0 & N/A \\
    \bottomrule
    \end{tabular}
    \caption{Jitter and skew distribution for each PDC and compared between the current and the previous revision. The skew corresponds to the variation of delay between each pixel. As the distributions are not properly normal, we quote both RMS (standard deviation) and full width at half maximum values.}
    \label{tab:timing}
\end{table}
The flag OR-tree contribution is estimated by subtracting in quadrature the simulated normalized propagation delay of the trigger tree from the measurement, since the simulated and measured skew for the combined flag tree and trigger tree are similar at \SI{\sim24}{\pico\second}.
In this case, the flag contribution would be around \SI{21.6}{\pico\second\rms}, an improvement over the \SI{37}{\pico\second\rms} observed in the first version. 
This reduction results from the improved matching of the flag H-tree in the current design.

\section{Adjustable hold-off time}\label{sec:mono}
\subsection{Material and methods}\label{subsec:met_HO}
The goal in this section is to compare the measured period of the hold-off to the simulation.
As presented in Section \ref{sec:arch}, each quenching circuit integrates 3 monostables for the hold-off, recharge, and flag duration.
To characterize the hold-off duration accurately, two measurement approaches are compared: a fast statistical method exploiting the digital sum histogram and a high-resolution technique employing an external time-to-digital converter.

The hold-off and recharge period is measured by keeping the FE\_TRIG high while the QCs are enabled and measuring the time interval of the resulting digital sum (\SI{10}{\nano\second} period). 
The height of the first bin represents the number of enabled QC. 
The time interval of peaks in the resulting digital sum acquisition represents the average hold-off and recharge period, while their distribution represents the pixel-to-pixel variation.

However, achieving a resolution below \SI{10}{\nano\second} on the monostable measurement requires an external time-to-digital converter, as shown in \RSfigtxt\ref{fig:block_flag}.
While keeping the trigger high and only enabling one pixel at a time, the Time Tagger is set to measure the period between two consecutive flags.
For a proper compromise between execution time and sufficient statistical representation, each pixel is triggered 32,000 times.
\subsection{Results}\label{subsec:res_HO}

\begin{figure}[!t]
    \centering
    \includegraphics[width=0.8\linewidth]{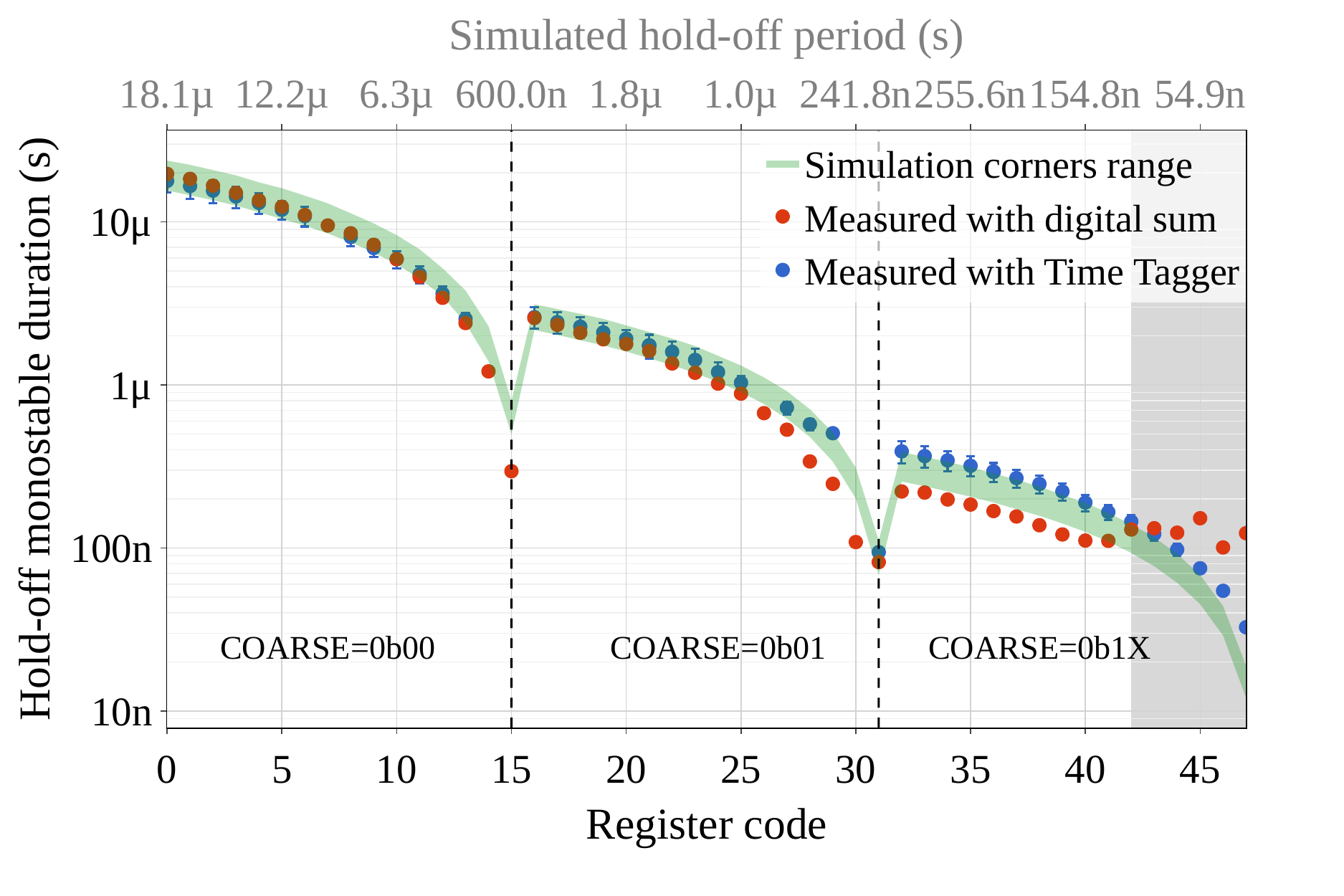} 
    \caption{Hold-off duration measured for each register code. The measured values are taken with either the Time Tagger setup or the digital sum setup. The gray zone corresponds to where the digital sum can't be used to estimate the hold-off. The simulated hold-off periods are labeled by their corresponding register codes, rather than as a continuous axis.}
    \label{fig:HO_var}
\end{figure}

\RSfigtxt\ref{fig:HO_var} presents the monostable duration measured as a function of register codes with both the digital sum and the time tagger, compared to the simulation corners.
Those simulation corners model the process variations inherent to the technology, covering nominal, fast-fast (FF), slow-slow (SS), fast-slow (FS) and slow-fast (SF) transistor variations.
\
The figure also shows the expected hold-off for each register code corresponding to each measured configuration.
These measurements provide a range of hold-off periods from \SI{32}{\nano\second} up to \SI{18}{\micro\second}.
Between registers 42 and 48, the gray zone shows where the digital sum resolution is inadequate to estimate the hold-off.
Furthermore, the three coarse configurations are intentionally overlapping at transition points to provide continuous hold-off period coverage across the entire range.
Since each coarse setting uses a different capacitance, the slope of the duration versus fine setting changes at each transition. 
When plotted on a logarithmic scale, these slope changes manifest as dips in \RSfigtxt\ref{fig:HO_var}.

The Time Tagger measurements follow the trend of the simulated values, with observed differences within fabrication process variations.
Besides, the smaller coarse setting shows a deviation from the simulated values, which limits the lower achievable period to around \SI{32}{\nano\second} instead of \SI{14}{\nano\second} due to the recharge period being included in the measured hold-off period.
\\

\subsection{Discussion}\label{subsec:dis_HO}
The method based on the digital sum reveals an average of  63\% decrease in the monostable duration, which is more noticeable in the fastest coarse setting. 
The increased digital activity caused by the enabled clock, the sampling of the digital sum, and all the pixels triggering at the same time affects these noise-susceptible analog circuits.
The expected simulation corner range comes from the simulation of a single quenching circuit, not integrated into an array, which might explain the period reduction.
This technique provides a good estimation of the hold-off delay while in operation without the use of external equipment.
The main drawback comes from the limited resolution dictated by the \SI{10}{\nano\second} period clock, which prevents the proper estimation of the hold-off delay below \SI{\sim80}{\nano\second}.

\section{Power consumption}\label{sec:power}
\subsection{Material and methods}\label{subsec:met_power}
The PDC's power consumption is categorized into two primary components, static and dynamic, over which come four different power supplies: 
\begin{itemize}
    \item 1V8 CORE: powers all digital circuits  and lower voltage of the quenching circuits (\SI{1.8}{\volt});
    \item 1V8 IO: supplies the output buffers (\SI{1.8}{\volt});
    \item 5V FE: powers the front-end part of the quenching circuit;
    \item 5V AM: supplies the analog monitor's current sources.
\end{itemize}\

\begin{figure}[!t]
    \centering 
    \includegraphics[width=0.75\columnwidth]{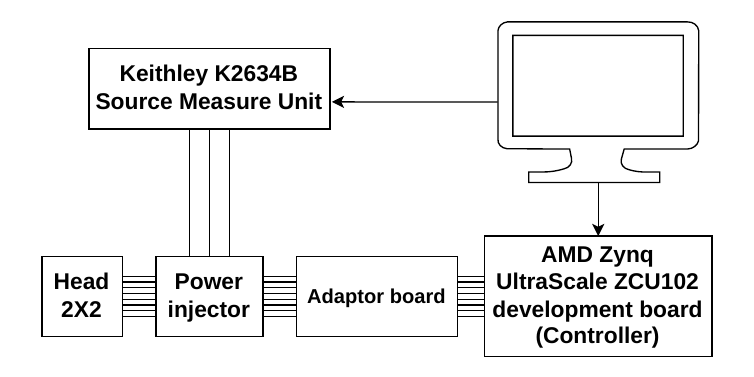} 
    \caption{Block diagram of the setup used to measure the power consumption of the ROIC. The tile controller communicates to the readout through the adaptor board and the power injector. The power injector enables communication with the ROIC while having switches to power the head board with the tile controller or an external power supply like the Keithley K2634B Source Measure Unit. }
    \label{fig:block_power}
\end{figure}

The power consumption of four PDCs is measured with a source measure unit (SMU) bypassing the adaptor board supplies through a power injector board (\RSfigtxt\ref{fig:block_power}).
The power injector feeds through the PDC and tile controller signals and provides input connectors for the external supplies from a Keithley K2634B SMU.\\
\
The power consumption analysis divides contributions from four sources: static power, QC events, clock activity, and data transmission. 
Static operation corresponds to the condition where the PDC is powered with no clock input or internal activity, remaining idle but ready to acquire.
Dynamic energy components (per event, per clock cycle, and per data transmission) are extracted by measuring power consumption with different rates and removing the static power offset.
Table \ref{tab:power_rate} presents the rate range used to characterize the dynamic power contribution of each circuit component, which is determined by the operational limits of the measurement platform.
\begin{table}[!t]
    \centering
    \begin{tabular}{lc}
        \toprule
        Component & Rate Range \\
        \midrule
        \midrule
        QC events & \SI{5}{\kilo\hertz} to \SI{5}{\mega\hertz} \\
        Clock & \SI{10}{\mega\hertz} to \SI{200}{\mega\hertz} \\
        Data transmission & \SI{100}{\mega\hertz} \\
        \bottomrule
    \end{tabular}
    \caption{Rate range for the characterization of the dynamic power contributions.}
    \label{tab:power_rate}
\end{table}

\subsection{Results}\label{subsec:res_power}
In static operation, the PDC consumes \SI{85(0.5)}{\micro\watt} on average on four PDCs.
The major part of this consumption comes from the polarization circuits for the monostables (current sources, operational amplifiers).

The QC power consumption from the transistors (\SI{5}{\volt} FE) follows a linear fit with the triggering frequency, presenting a regression score ($R^2$) of \SI{0.9998}{} on all the ranges specified in Table \ref{tab:power_rate}.
In total, each trigger per pixel consumes \SI{35}{\pico\joule}, \SI{25}{\pico\joule} of which comes from the front-end part of the QC (5V FE).
The power consumption ($P_{events}$) based on the SPAD trigger rate ($R_{SPAD}$, expressed in counts per second, cps) is determined by \eqref{eq:power_event}.
\begin{equation}\label{eq:power_event}
    P_{events} = 35\times10^{-12} (J) \times R_{SPAD}
\end{equation}

The power consumption of the 1V8 CORE supply with the clock activated follows a linear trend ($R^2=0.998$) over the entire measurement range, corresponding to \SI{0.8}{\nano\joule} per clock cycle. 
At the nominal clock frequency of \SI{100}{\mega\hertz}, this results in \SI{80}{\milli\watt} when the system samples continuously the digital sum.
The power consumption coming from the clock rate follows \eqref{eq:power_clock}, which approximates the power consumption ($P_{clock}$) based on the number of samples from the digital sum per event ($N_{clock \text{ } cycles/acq}$) and the average event acquisition rate ($R_{acq}$).
\begin{equation}\label{eq:power_clock}
    P_{clock} =0.8\times10^{-9} (J)\times R_{acq}\times N_{clock \text{ } cycles/acq}
\end{equation}

The power consumption from the 1V8 IO output buffers consumes \SI{79}{\pico\joule} per transition when transmitting back the configuration clock at \SI{100}{\mega\hertz}.
This results in a power consumption of \SI{7.89}{\milli\watt} on the IO supply.
The consumption depends on the parasitic capacitance driven by the buffer, which in this case is a cable of around \SI{30}{\centi\meter}.
Since the transmission is also on request, \eqref{eq:power_buf} provides an approximation of the buffer's power consumption ($P_{buf}$) based on the amount of transmitted data ($N_{bits/acq}$).
\begin{equation}\label{eq:power_buf}
    P_{buf} =79\times 10^{-12} (J) \times R_{acq} \times N_{bits/acq}
\end{equation}

\subsection{Discussion}\label{subsec:dis_power}
\begin{figure}[!t]
    \centering
    \includegraphics[width=0.8\linewidth]{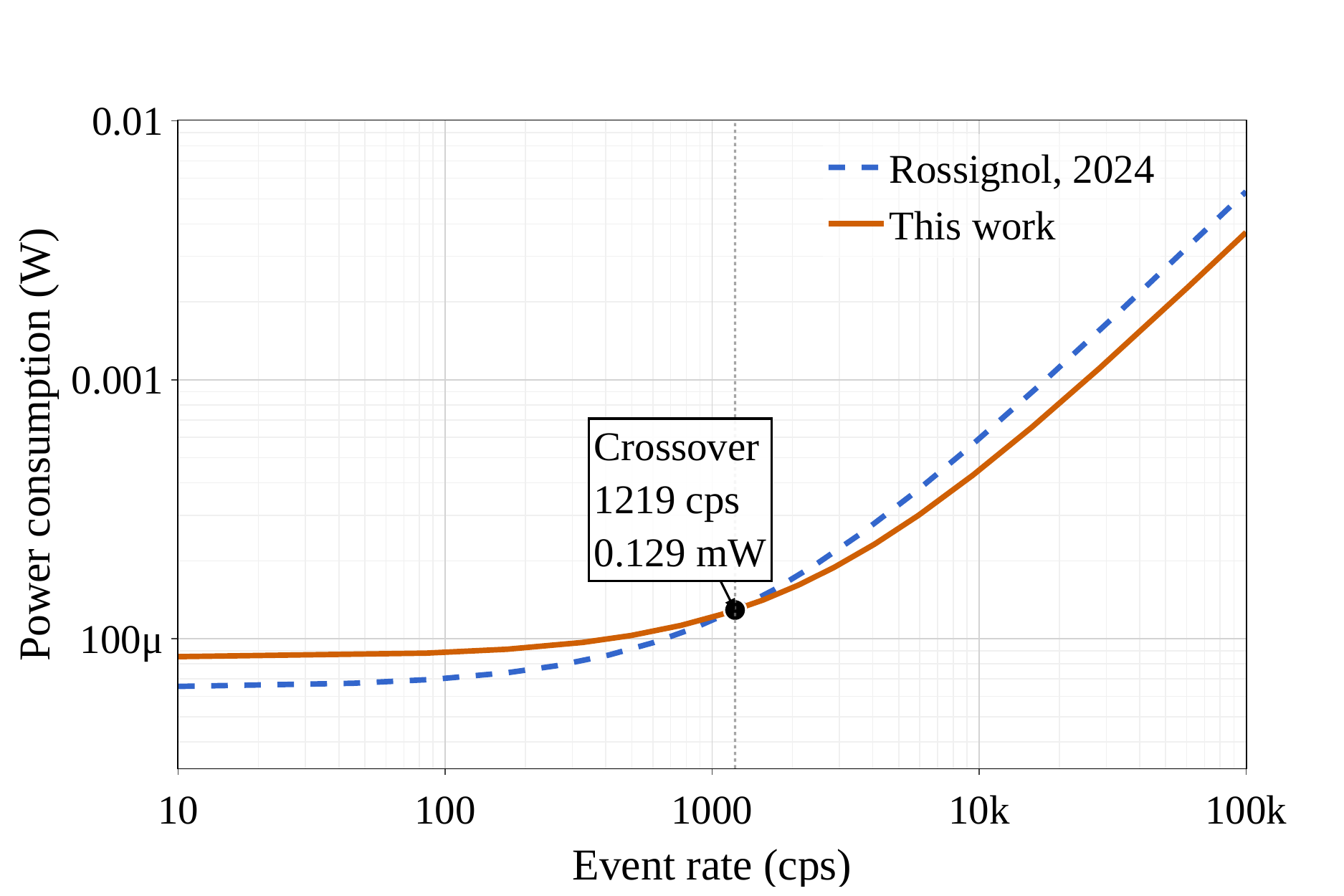} 
    \caption{Comparison of power consumption as a function of event rate between \cite{rossignol3DPhotontoDigitalConverter2024} and the current work for all supplies summed together. Below the crossover point (\SI{1219}{cps}), \cite{rossignol3DPhotontoDigitalConverter2024} exhibits lower power consumption, while the current work is more efficient at higher event rates.}
    \label{fig:power_comp}
\end{figure}

The power consumption is impacted by the photon rate and the sampling parameters represented by equations (\ref{eq:power_event}), (\ref{eq:power_clock}), and (\ref{eq:power_buf}).
The energy consumed per clock cycle was reduced from \SI{1.8}{\nano\joule} in \cite{rossignol3DPhotontoDigitalConverter2024} to \SI{0.8}{\nano\joule} in the proposed architecture, corresponding to a 2.25-fold improvement.
This reduction is primarily achieved through the introduction of clock-gating techniques that reduce unnecessary switching activity in the digital logic and clock distribution network. 
Since clock distribution represents the dominant contributor to the digital power consumption, the proposed gating strategy significantly improves energy efficiency, particularly in low-event-rate operating conditions.
\RSfigtxt\ref{fig:power_comp} shows the power consumption from all supplies of the first revision compared to the current revision based on the photon rate.
\

The increase in static power consumption from \SI{65}{\micro\watt} to \SI{85}{\micro\watt} causes this revision to consume more power at event rates below \SI{1.2}{\kilo\hertz}, as shown in \RSfigtxt\ref{fig:power_comp}. 
This increase primarily results from the addition of bias stabilization circuits for the monostables, which were introduced to improve monostable timing stability across temperature and process variations.
However, beyond \SI{10}{\kilo\hertz}, this revision achieves approximately \SI{30}{\percent} lower total power consumption, making it significantly more efficient at higher event rates.
Note that the measured power consumption comprises the energy dissipated by output buffers driving \SI{30}{\centi\meter} cables. 
When integrated into tiles, the interconnection length will be reduced to approximately \SI{5}{\centi\meter}, decreasing the capacitive load by a factor of six and correspondingly reducing the power consumption of the output buffers.

\section{Conclusion}\label{sec:conclusion}

The current work presents an improved revision of the PDC ROIC presented in \cite{rossignol3DPhotontoDigitalConverter2024}.
This ROIC is designed for 3D-integration of a \qtyproduct{5x5}{\milli\meter} array of 4096 individually quenched SPADs.
The readout comprises a digital sum of the triggered SPADs, programmable outputs for flexibility purposes, and a trigger tree propagated across the array for debug purposes.
This whole architecture is developed for low-power applications but also for sub-\SI{100}{\pico\second} precision in meter-scale systems.
With the redesign of the whole trigger tree, the PDC flag can be more accurately characterized, which gives an overall jitter of \SI{\sim21.6}{\pico\second\rms}.
Moreover, the tunable hold-off period, ranging from \SI{32}{\nano\second} up to \SI{18}{\micro\second}, allows for effective afterpulsing mitigation by accommodating the temperature-dependent carrier trapping dynamics across a broad operating range. 
Additionally, the digital sum facilitates rapid chip characterization by providing a way to estimate the hold-off duration on almost all the range.
With the current revision, the readout sees a \SI{30}{\percent} decrease in power consumption at event rates above \SI{10}{\kilo\hertz}, while observing a \SI{20}{\percent} increase in static power consumption.
Having characterized the ROIC itself, we will now focus on testing complete PDCs, i.e., with 3D SPADs, and compare the same performance parameters.
We will also gather performances of more devices to provide insight	on their dispersion, which is needed to inform system designs.
This will support the assembly of PDM with multiple PDCs for integration in meter-scale applications.

\acknowledgments
The present work was supported in part by the United States Department of Energy, Office of Defense Nuclear Nonproliferation Research and Development in the National Nuclear Security Administration.
The authors also acknowledge the financial contributions of the Natural Sciences and Engineering Research Council of Canada (CRSNG), the Fonds de recherche du Québec - Nature et technologies (FRQNT), the Arthur B. McDonald Canadian Astroparticle Physics Research Institute (Queen's~U.), the Regroupement strat\'egique en microsyst\`emes du Qu\'ebec (ReSMiQ), and the Canada Foundation for Innovation (CFI).
AI (Claude and ChatGPT) was utilized to improve the linguistic precision and clarity of the technical writing throughout this article.




\bibliographystyle{JHEP}
\bibliography{references}
\end{document}